# Dependence of tunnel magnetoresistance in MgO based magnetic tunnel junctions on Ar pressure during MgO sputtering

Shoji IKEDA[1], Jun HAYAKAWA[1,2], Young Min LEE[1], Ryutaro SASAKI[1], Toshiyasu MEGURO[1], Fumihiro MATSUKURA[1] and Hideo OHNO[1]

*1 Laboratory for Nanoelectronics and Spintronics, Research Institute of Electrical Communication, Tohoku University, 2-1-1 Katahira, Aoba-ku, Sendai 980-8577, Japan*

*2 Hitachi Ltd., Advanced Research Laboratory, 1-280, Higashi-koigakubo, Kokubunji-shi, Tokyo 185-8601, Japan*

We investigated dependence of tunnel magnetoresistance effect in CoFeB/MgO/CoFeB magnetic tunnel junctions on Ar pressure during MgO-barrier sputtering. Sputter deposition of MgO-barrier at high Ar pressure of 10 mTorr resulted in smooth surface and highly (001) oriented MgO. Using this MgO as a tunnel barrier, tunnel magnetoresistance (TMR) ratio as high as 355% at room temperature (578% at 5K) was realized after annealing at 325 $^{o}$C or higher, which appears to be related to a highly (001) oriented CoFeB texture promoted by the smooth and highly oriented MgO. Electron-beam lithography defined deep-submicron MTJs having a low-resistivity Au underlayer with the high-pressure deposited MgO showed high TMR ratio at low resistance-area product (*RA*) below 10 $\Omega\mu m^2$ as 27% at *RA* = 0.8 $\Omega\mu m^2$, 77% at *RA* = 1.1 $\Omega\mu m^2$, 130% at *RA* = 1.7 $\Omega\mu m^2$, and 165% at *RA* = 2.9 $\Omega\mu m^2$.

E-mail: sikeda@riec.tohoku.ac.jp

Development of magnetic tunnel junctions (MTJs) that exhibit high tunnel magnetoresistance (TMR) ratio is needed to further increase the capacity of magnetoresistive-random-access-memories (MRAMs), and high TMR at low resistance area ($RA$) product range is required for magnetic sensors in the sub Tbit/in$^2$ hard-disk-drives and beyond. Conventional MTJs with amorphous aluminum oxide barriers yield TMR ratio ranging 18–70%,[1-3] which is not high enough for these applications. Recent developments of MTJs based on highly ordered (001)-oriented Fe(Co) electrodes with a MgO barrier, motivated by pioneering theoretical predictions,[4-6] resulted in very high TMR ratio ranging from 67 % to 260% at room temperature (RT) with $RA$ products spanning from a few ten $\Omega\mu m^2$ and higher.[7-13] Very recently, high TMR ratio of up to 138% was reported at low $RA$ of 2.4 $\Omega\mu m^2$ using a CoFeB/ultrathin-Mg/MgO/CoFeB structure,[14, 15] where an ultrathin-Mg was inserted to maintain highly (001)-oriented MgO at the very thin MgO thickness level required for achieving the low $RA$. These results have suggested that the (001) oriented state of MgO is related to the TMR ratio. Because it has been known that the film structure such as morphology and orientation depends on Ar pressure during sputtering, [16] we have investigated the dependence of the quality of MgO on the Ar pressure during MgO sputtering. Here, we show that one can obtain giant TMR ratio as high as 355 % at $RA$ of 547 $\Omega\mu m^2$ by selecting the pressure. We also show that high TMR ratio and low $RA$, without inserting ultrathin-Mg, can be realized using this high pressure deposition of MgO combined with a low resistivity underlayer. [17)]

The MTJ films used in this study were deposited onto a thermally oxidized Si(001) wafer using a magnetron sputtering system with a base pressure of less than $10^{-9}$ Torr. The MTJ stack structure consists of substrate/ Ta(5)/ underlayer(50)/ Ta(5)/ NiFe(5)/ MnIr(8)/ CoFe(2)/ Ru(0.8)/ $Co_{40}Fe_{40}B_{20}$ (3)/ MgO($t_{MgO}$)/ $Co_{40}Fe_{40}B_{20}$ (3)/ Ta(5)/ Ru(15) (in nm), where Ru and Au were used as the underlayer, and nominal MgO thickness $t_{MgO}$ was varied from 0.8 nm to 3.0 nm. The MgO barrier was sputtered at the Ar gas pressure in the range of 1 to 20 mTorr (deposition rate range of 0.014-0.020 nm/s). When the pressure was too low (below 0.2 mTorr) or too high (above 30mTorr) the discharge became unstable. The Ar pressure for metal sputtering was 1 mTorr. MTJs with a junction area of 0.8 x 4.0 $\mu m^2$ were prepared by photolithography, whereas those with an area of 80 x 160 $nm^2$ were fabricated using electron-beam lithography. The patterned MTJs were annealed for 1 h at each prescribed temperature in the range from 250 $^oC$ to 425 $^oC$ in a vacuum of $10^{-6}$ Torr under a magnetic field of 4 kOe. The electrical properties of MTJs were measured at RT and at 5K using a dc four probe method. The TMR ratio was calculated as $(R_{AP}-R_P)$ /$R_P$ x 100, where $R_{AP}$ and $R_P$ are the resistance for antiparallel (AP) and parallel (P) magnetization configurations between the free and reference layers. The film structure and the surface morphology were characterized by x-ray diffraction (XRD) using Cu-K$_\alpha$ radiation and atomic force microscope (AFM), respectively.

TMR ratios as a function of annealing temperature ($T_a$) for MTJs with an 1.5 nm-thick MgO barrier deposited under three different Ar pressures are shown in Fig. 1. When annealed at $T_a$ of less than 300$^o$C, the TMR ratios of the three series of MTJs are virtually the same. The effect of Ar pressure starts to appear at $T_a$ higher than 325$^o$C. The maximum TMR ratios of the three series are 260% ($T_a$ =375$^o$C) for 1 mTorr, 289% ($T_a$ =375$^o$C) for 3 mTorr, and 355% ($T_a$ =400$^o$C) for 10 mTorr. According to Julliere's formula,[18] these TMR ratios correspond to tunneling spin polarizations of 0.75, 0.77 and 0.80, respectively. The barrier height ($\phi$) deduced from $RA$ vs. $t_{MgO}$ curves (not shown) is about 0.35 eV regardless of the Ar pressure, showing that the TMR ratio does not critically depend on the pressure nor on $T_a$ in accordance with previous reports.[9, 13, 14] These observations indicate that the difference of the improved MgO quality at high Ar pressure described later does not have direct impact on TMR ratio, because no clear difference of the TMR ratio is observed at $T_a$ below 300$^o$C. The fact that the effect of MgO sputtering pressure becomes obvious at $T_a$ above 325$^o$C suggests that the quality difference of MgO barrier depending on sputtering pressure may be affecting the crystallization of the amorphous CoFeB free and reference electrodes, because it is known that CoFeB electrodes start to form (001) bcc textures at $T_a = 325^oC$,[13, 19] *i.e.*, the higher crystalline orientation and smoother surface of MgO is likely to lead to highly (001) oriented CoFeB texture, making the structure closer to the ideal epitaxial one that is assumed by the theories[4, 5] leading to an enhanced TMR ratio.

Figure 2 shows the maximum TMR ratio obtained from MTJs annealed at $T_a$ ranging from $325^oC$ to $400\,^oC$ as a function of $RA$ at parallel magnetization configuration. For $RA > 10\ \Omega\mu m^2$, the junction size of 0.8 x 4.0 $\mu m^2$ was used; below $10\ \Omega\mu m^2$, the junction size was chosen to be 80 x 160 $nm^2$, in order to minimize the effect of the electrode resistance. When the MTJs with Ru underlayer are compared, Fig. 2 indicates that TMR ratio increases with increasing sputtering pressure and then saturates. By adopting low-resistivity Au for underlayer with a MgO barrier deposited at 10 mTorr, high TMR ratio (filled triangles) is obtained in the lower $RA$ range from 0.8 $\Omega\mu m^2$ to 10 $\Omega\mu m^2$. Typical TMR ratio of the MTJs with Au underlayer is 27% at $RA = 0.8\ \Omega\mu m^2$ and 77% at $RA = 1.1\ \Omega\mu m^2$, which is four to ten times greater than the ratios of current perpendicular to plane giant magnetoresistance spin valve films having similar $RA$.[20, 21] We also obtained TMR ratios of 130% at $RA = 1.7\ \Omega\mu m^2$ and 165% at

$RA = 2.9$ $\Omega\mu m^2$, which are about eight to ten times greater than the ratios of aluminum

oxide barrier MTJs with similar $RA$.[22, 23]

Figure 3 shows the bias voltage dependence of TMR ratio at 5K and RT for a

MTJ with a 1.5 nm-thick MgO barrier deposited at 10 mTorr and annealed at 400$^o$C.

This device has a Ru underlayer. The bias voltage was defined as positive when

electrons were flowing from the bottom to the top layer. By lowering the measurement

temperature from RT to 5K, TMR ratio increased from 355% to 578% (equivalent to

spin polarization of 0.86). As can be seen from the figure, TMR ratio decreased with

increasing the bias voltage. The bias voltages ($V_{half}$), where the TMR ratio drops half of

its maximum, were -0.54V and +0.59V at RT, and -0.29V and +0.34V at 5K. For the

curve at 5K, a pronounced decrease of TMR ratio can be seen in the low bias range of

±0.5V as compared with the curve at RT. For both curves at RT and 5K, $V_{half}$ at positive

bias is slightly higher (50 mV) than that at negative bias, which is much less than what

was reported in fully epitaxial Fe/MgO/ Fe MTJs.[9, 12]

To investigate the orientation and surface morphology of MgO barriers, XRD

($2\theta$-$\theta$ scan and rocking curve measurement) and AFM measurements were performed.

Figure 4(a) shows the XRD $2\theta$-$\theta$ patterns for two structures, substrate/ Ta(5)/

$Co_{40}Fe_{40}B_{20}(3)$/ MgO ($t_{MgO}$=100) structure, where the MgO layers were deposited at 1

mTorr (black line) and 10 mTorr (gray line). Although the MgO layers of these samples were thicker than the actual MgO barrier in the MTJs, we believe that it still provides valuable information on the morphology of MgO barrier. With increasing the Ar pressure during MgO deposition, the intensity of MgO (002) diffraction peak increased, as can be seen in Fig.4 (a). Similar increase was also observed in the MgO (004) diffraction peak (not shown). Fig. 4(b) shows the rocking curves of the two MgO films at (002). The full width at half maximum (FWHM) is about $5.1^o$ for the MgO layer deposited at 1 mTorr, and $4.4^o$ for the MgO layer deposited at 10 mTorr, *i.e.* high Ar pressure reduced FWHM. The average roughness ($R_a$) of the top surface of substrate/ Ta(5)/ $Co_{40}Fe_{40}B_{20}$(3)/ MgO($t_{MgO}$) structures having different $t_{MgO}$ evaluated from AFM images is compiled in Fig. 5. The high-pressure deposition resulted in low $R_a$ in all the $t_{MgO}$ range investigated (from 5 nm to 100 nm). Similar XRD profiles and AFM images were obtained after annealing at $375^oC$, showing that these structural features of MgO barrier were maintained after annealing. Although the mechanism responsible for the structural improvement of MgO at high Ar pressure is not understood at the moment, this structural improvement is believed to be responsible for the enhancement of TMR ratio after annealing above the crystallization temperature of amorphous CoFeB electrodes as discussed earlier. We note that TMR ratio as high as 355% is obtained in this work, even though FWHM of the rocking curve is still broad compared to MgO single crystals (FWHM=0.3$^o$-0.5$^o$ in commercial products), which suggests that there may still be room for further increasing TMR ratio by increasing the (001) orientation of CoFeB crystallized texture and of MgO barrier.

In summary, we investigated the influence of the MgO-barrier sputtering pressure on the TMR ratio for MTJs. The deposition of MgO barrier at Ar pressure of 10 mTorr resulted in highly (001) oriented MgO with smooth surface with which we obtained the TMR ratio as high as 355% at RT (578% at 5K) at $RA$ of 547 $\Omega\mu m^2$ after annealing at 325 $^o$C or higher. It is suggested that the high TMR ratios are related to the highly (001) oriented CoFeB texture promoted upon annealing by the smooth and highly oriented MgO. By the use of low-resistivity Au underlayer and the deep-submicron junctions (size of 80 x 160 nm$^2$) fabricated using electron-beam lithography, in addition to high-pressure deposition of MgO barrier, we obtained the TMR ratio in the low $RA$ regime as 27% at $RA$ = 0.8 $\Omega\mu m^2$, 77% at $RA$ = 1.1 $\Omega\mu m^2$, 130% at $RA$ = 1.7 $\Omega\mu m^2$, and 165% at $RA$ = 2.9 $\Omega\mu m^2$.

This work was supported by the IT-program of Research Revolution 2002 (RR2002): "Development of Universal Low-power Spin Memory" from the Ministry of Education, Culture, Sports, Science and Technology of Japan.

References

1) T. Miyazaki and N. Tezuka: J. Magn. Magn. Mater. **139** (1995) L231.

2) J. S. Moodera, L. R. Kinder, T. M. Wong and R. Meservey: Phys. Rev. Lett. **74** (1995) 3273.

3) D. Wang, C. Nordman, J. Daughton, Z. Qian, and J. Fink: IEEE Trans. Magn. **40** (2004) 2269.

4) W. H. Butler, X.-G. Zhang, T. C. Schulthess, and J. M. MacLaren: Phys. Rev. B **63** (2001) 054416.

5) J. Mathon and A. Umersky: Phys. Rev. B **63** (2001) 220403R.

6) X.-G. Zhang and W. H. Butler: Phys. Rev. B **70** (2004) 172407.

7) J. Faure-Vincent, C. Tiusan, E. Jouguelet, F. Canet, M. Sajieddine, C. Bellouard, E. Popova, M. Hehn, F. Montaigne, and A. Schuhl: Appl. Phys. Lett. **82** (2003) 4507.

8) S. Yuasa, A. Fukushima, T. Nagahama , K. Ando, and Y. Suzuki: Jpn. J. Appl. Phys. **43** (2004) L588.

9) S. Yuasa, T. Nagahama, A. Fukushima, Y. Suzuki, and K. Ando: Nat. Mater. **3** (2004) 868.

10) S. S. Parkin, C. Kaiser, A. Panchula, P. M. Rice, B. Hughes, M. Samant, and S.-H. Yang: Nat. Mater. **3** (2004) 862.

11) D. D. Djayaprawira, K. Tsunekawa, M. Nagai, H. Maehara, S. Yamagata, N. Watanabe, S. Yuasa, Y. Suziki, and K. Ando: Appl. Phys. Lett. **86** (2005) 092502.

12) T. Nozaki, A. Hirohata, T. Tezuka, S. Sugimoto, and K. Inomata: Appl. Phys. Lett. **86** (2005) 082501.

13) J. Hayakawa, S. Ikeda, F. Matsukura, H.Takahashi, and H. Ohno: Jpn. J. Appl. Phys. **44** (2005) L587.

14) K. Tsunekawa, D. D. Djayaprawira, M. Nagai, H. Maehara, S. Yamagata, N. Watanabe, S. Yuasa, Y. Suzuki, and K. Ando: Appl. Phys. Lett. **87** (2005) 072503.

15) H. Kubota, A. Fukushima, Y. Ootani, S. Yuasa, K. Ando, H. Maehara, K. Tsunekawa, D. D. Djayaprawira, N. Watanabe and Y. Suzuki: Jpn. J. Appl. Phys. **44** (2005) L1237.

16) J. A. Thornton: J. Vac. Sci. Technol., 12, (1974) 830.

17) J. Hayakawa, S. Ikeda, Y. M. Lee, R. Sasaki, T. Meguro, F. Matsukura, H.Takahashi, and H. Ohno: Jpn. J. Appl. Phys. **44** (2005) L1267.

18) M. Julliere: Phys. Lett. **54A** (1975) 225.

19) S. Cardoso, C. Cavaco, R. Ferreira, L. Pereira, M. Rickart, P. P. Freitas, N. Franco, J. Gouveia, and N. P. Barradas: J. Appl. Phys. **97** (2005) 10C916.

20) A. Tanaka, Y. Seyama, A. Jogo, H. Oshima, R. Kondo, H. Kishi, C. Kamata, Y.

Shimizu, S. Eguchi, K. Satoh: IEEE Trans. Magn. **40** (2004) 203.

21) H. Fukuzawa, H. Yuasa, S. Hashimoto H. Iwasaki, and Y. Tanaka: Appl. Phys. Lett. **87** (2005) 082507.

22) S. Mao, E. Linville, J. Nowak, Z. Zhang, S. Chen, B. Karr, P. Anderson, M. Ostrowski, T. Boonstra, H. Cho, O. Heinonen, M. Kief, S. Xue, J. Price, A. Shukh, N. Amin, P. Kolbo, P-L. Lu, P. Steiner, Y.C. Feng, N-H. Yeh, B. Swanson, and P. Ryan: IEEE Trans. Magn. **40** (2004) 307.

23) T. Kuwashima, K. Fukuda, H. Kiyono, K. Sato, T. Kagami, S. Saruki, T. Uesugi, N. Kasahara, N. Ohta, K. Nagai, N. Hachisuka, N. Takahashi, M. Naoe, S. Miura, K. Barada, T. Kanaya, K. Inage, and A. Kobayashi: IEEE Trans. Magn. **40** (2004) 176.

Figure caption

Fig. 1. TMR ratios as functions of annealing temperature ($T_a$) for MTJs with an 1.5 nm-thick MgO barrier deposited under three different Ar pressures (1, 3 and 10 mTorr).

Fig. 2. TMR ratios as functions of $RA$ at RT in parallel magnetization configuration for the Ru underlayer MTJs with the MgO barrier sputtered at the pressure of 1, 3, 10 and 20 mTorr (open symbols), and for the Au underlayer MTJs with the MgO barrier sputtered at 10 mTorr (filled triangles). These MTJs were annealed at optimal temperature showing the highest TMR ratio in the temperature range from 325$^o$C to 400$^o$C.

Fig. 3. Bias voltage dependence of the TMR ratio at 5K and RT for a MTJ, which was annealed at 400$^o$C for 1hr, with an 1.5 nm-thick MgO barrier sputtered at 10 mTorr.

Fig. 4. (a) XRD $2\theta$-$\theta$ patterns and (b) rocking curves of MgO(002) diffraction peak for the substrate/ Ta(5)/ $Co_{40}Fe_{40}B_{20}$(3)/ MgO(100) structure (in nm), where the MgO layers were deposited at 1 mTorr and 10 mTorr.

Fig. 5. Average roughness ($R_a$) of the top surface of substrate/ Ta(5)/ $Co_{40}Fe_{40}B_{20}(3)$/ MgO($t_{MgO}$) structures having different $t_{MgO}$ evaluated from AFM images, where the MgO layers were sputtered at 1 mTorr and 10 mTorr.

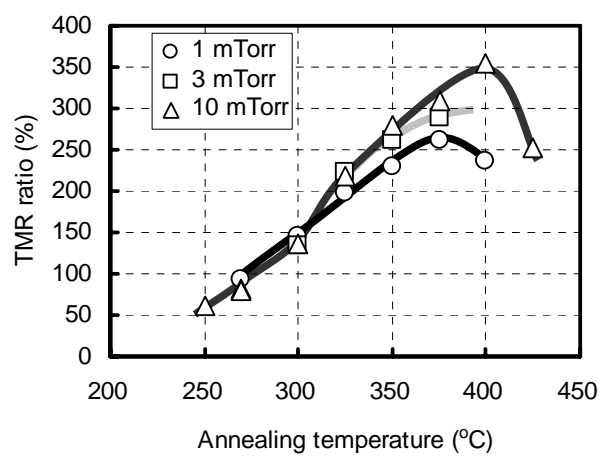

Fig. 1

S.Ikeda et. al.

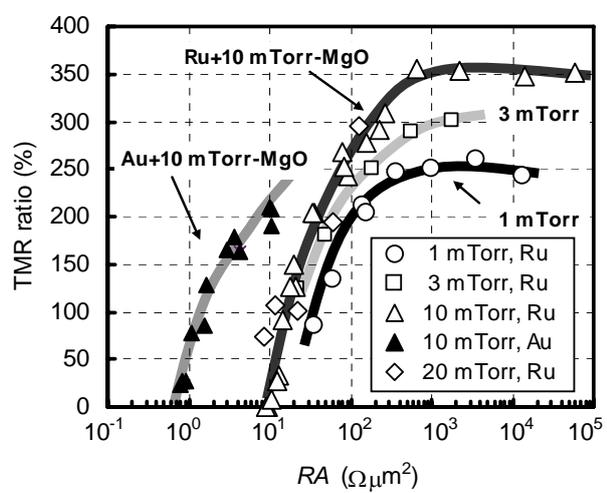

Fig. 2

S.Ikeda et. al.

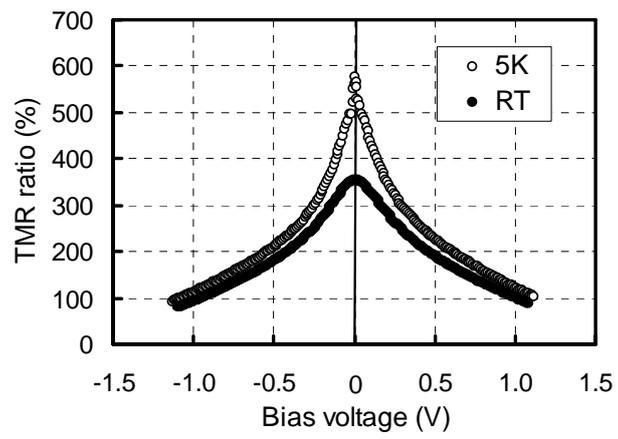

Fig. 3

S.Ikeda et. al.

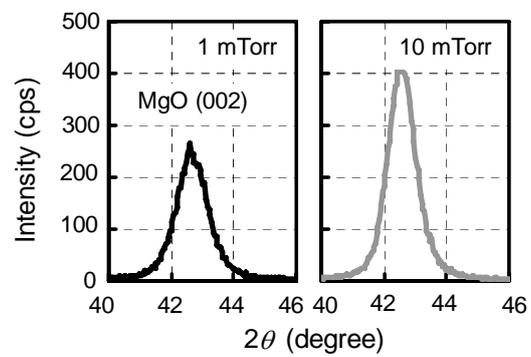

(a)

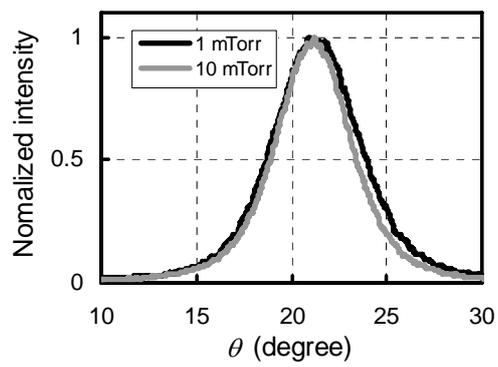

(b)

Fig.4

S.Ikeda et. al.

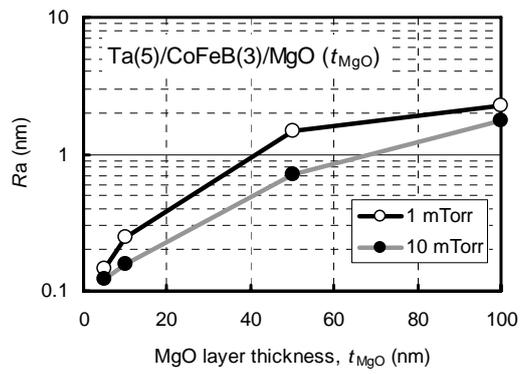

Fig.5

S.Ikeda et. al.